\documentclass[12pt]{article}
\pdfoutput=1
\usepackage{amsmath,amssymb,amsfonts,amsthm,bm,bbm,cancel,wasysym}
\usepackage{epsfig,graphics,graphicx,epstopdf,caption,subcaption}
\usepackage[numbers,sort&compress]{natbib}
\graphicspath{{Charts/}}
\usepackage{array,booktabs,colortbl,colordvi,multirow}
\usepackage{colordvi,color,xcolor}
\usepackage{hyperref}
\usepackage{rotating}
\usepackage{comment}
\usepackage{feynmf}

\begin{document}

\begin{center}

{\Large {\bf R-Symmetric NMSSM}}\\

\vspace*{0.75cm}

{Shuai Xu and Sibo Zheng}

\vspace{0.5cm}
{Department of Physics, Chongqing University, Chongqing 401331, China}

\end{center}
\vspace{.5cm}

\begin{abstract}
 \noindent 
It is well known that the observed Higgs mass is more naturally explained in the NMSSM than in the MSSM.
Without any violation of this success, there are variants on the NMSSM which can lead to new phenomenologies. 
In this study we propose a new variant of NMSSM by imposing an unbroken $R$ symmetry.
We firstly identify the minimal structure of such scenario from the perspective of both simplicity and viability,
then compare model predictions to current experimental limits, 
and finally highlight main features that differ from the well-known scenarios.
\end{abstract}

\renewcommand{\thefootnote}{\arabic{footnote}}
\setcounter{footnote}{0}
\thispagestyle{empty}
\vfill
\newpage
\setcounter{page}{1}

\section{Introduction}
In the context of supersymmetry (SUSY), 
the Minimal Supersymmetric Standard Model (MSSM) is not favored by the observed Higgs mass \cite{1207.7214,1207.7235} if naturalness is taken into account. 
The naturalness problem can be relieved by an addition of  Standard Model (SM) singlet which directly couples to the two Higgs doublets in the MSSM.
Compared to the MSSM, in this singlet extended model known as the Next-to-Minimal Supersymmetric Standard Model (NMSSM),
the observed Higgs mass \cite{1112.2703,1201.2671,1202.5821,1108.1284,0006198,1005.1282}  is easily explained  in terms of new tree-level contribution instead of large loop corrections.
Apart from the Higgs mass issue, the conventional NMSSM has distinctive 
features such as a relaxed constraint on the neutralino dark matter from dark matter direct detections \cite{1805.12562,1708.06917,1705.03380},  
and even enables us to address baryon asymmetry by the means of strong first-order phase transition \cite{0606298, 1110.4378,1407.4134}, among others.

In fact, without any violation of the success on the Higgs mass naturalness, 
the conventional NMSSM can be modified in few different ways where new phenomenological outcomes 
may be more favored by current experiments. 
Motivated by earlier studies based on $R$ symmetry \cite{0206102, 0712.2039},
in this study we focus on the NMSSM with an unbroken $U(1)_R$ global symmetry, which will be referred as RNMSSM.
Because of $U(1)_R$  symmetry, 
we will show that the Kahler potential and the superpotential in the RNMSSM have to change as
\begin{eqnarray}
K&=&K_{\rm{MSSM}} + N^{\dag}N+\delta K, \label{Kahler} \\
W&=&W_{\rm{MSSM}}+\lambda NH_{u}H_{d}+\delta W,\label{W}
\end{eqnarray}
where $K_{\rm{MSSM}}$ and $W_{\rm{MSSM}}$ denote the Kahler potential and superpotential in the MSSM respectively,
$N$ is the singlet, and $H_{u}$, $H_d$ are two Higgs doublets.
Accordingly, we expect soft terms in $\mathcal{L}_{\rm{soft}}$  to be altered as well.

The main goal of this study is to uncover the explicit forms of $\delta K$ and $\delta W$ from perspective of both simplicity and viability,
where ``simplicity" means that we seek a minimal extension on the matter content, while ``viability" requires that the RNMSSM should be consistent with current experimental bounds.
We organize the rest of materials as follows.
In Sec.II, we highlight the $R$ charge assignments in order to accomodate the $U(1)_R$  symmetry.
In Sec.III, we firstly infer the structure of the minimal RNMSSM, 
then discuss the main features of the supersymmetric mass spectrum and phenomenological implications from gauginos, neutralinos, sfermions to goldstino.
Finally, we conclude in Sec.IV.

\section{R Symmetry}
To be consistent with the $U(1)_R$ symmetry, the $R$ charges of some matter fields should be assigned as what follows. 
\begin{itemize}
\item The $R$ charges of Higgs doublets have to be zero in order to ensure an unbroken $R$ symmetry after the electroweak symmetry breaking.
This R charge assignment makes our model different from ref.\cite{0206102}, where the $R$ symmetry is broken by the vacuum expectation values (vevs) of the Higgs doublets.
\item The $R$ charge of $N$ should be two from the Yukawa interaction $W\supset NH_{u}H_{d}$ in Eq.(\ref{W}), 
and the vev of this scalar field should be zero as well, i.e, $\left<N\right>=0$.
\item The $R$ charges of SM gauge superfields $W^{i}_{\alpha}$  ($i=1-3$) have to be one\footnote{Spinor $\theta$ has mass dimension $-1/2$ and $R$ charge $+1$. So, for a chiral superfield with $R$ charge $q$, the scalar and fermion  field have $R$ charge $q$ and $q-1$, respectively. Likewise, for spinor gauge superfield with $R$ charge $q$, the fermion and gauge  field have $R$ charge $q$ and $q-1$, respectively.}, which make sure that all of the SM gauge fields are neutral under $U(1)_R$.
\end{itemize}
From the $R$ charge assignments above, one can infer the $R$ charges of all matter superfields from Eq.(\ref{W}), 
which are explicitly shown in Table.\ref{minimal}. 
They imply that all of SM fields do not carry the $R$ charge. 

\begin{table}
\begin{center}
\begin{tabular}{ccc}
\hline\hline
$\rm{Field}$~~~~ & $(\rm{SU}(3)_{c}, \rm{SU}(2)_{L})_{\rm{U(1)_{Y}}}$~~~ & $\rm{U}(1)_{\rm{R}}$ \\ \hline
$Q$~~ &  $(3,2)_{1/6}$~~~~ & $1$ \\
$\bar{u}$~~ &  $(\bar{3},1)_{-2/3}$ ~~~~& $1$  \\
$\bar{d}$~~ &  $(\bar{3},1)_{1/3}$ ~~~~& $1$  \\
$L$~~ &  $(1,2)_{-1/2}$~~~~ & $1$ \\
$\bar{e}$~~ &  $(1,1)_{1}$ ~~~~& $1$  \\  \hline
$H_{u}$~~ &  $(1,2)_{1/2}$~~~~ & $0$ \\
$H_{d}$~~ &  $(1,1)_{-1/2}$ ~~~~& $0$  \\
\hline
 $N$~~ &  $(1,1)_{0}$~~~~ & $2$  \\
$\sigma_{1}$~~ &  $(1, 1)_{0}$ ~~~~& $0$ \\
$\sigma_{3}$~~ &  $(1, 3)_{0}$ ~~~~& $0$  \\
$T$~~ &  $(1,3)_{0}$ ~~~~& $2$ \\
$\sigma_{8}$~~ &  $(8,1)_{0}$ ~~~~& $0$ \\
\hline \hline
\end{tabular}
\caption{Matter content and $R$ charge assignments in the minimal RNMSSM, where $\sigma_1$, $\sigma_{3}$, $T$, and $\sigma_8$ refer to 
an $SU(2)_L$ singlet, two $SU(2)_L$ triplets and  an $SU(3)_c$ octet states, respectively. 
The triplets (octet) make the charginos and neutralinos (gluino) Dirac fermions, see details in Sec.III.}
\label{minimal}
\end{center}
\end{table}

\section{Phenomenology of The Minimal RNMSSM}
In this section, we discuss the phenomenological features in the minimal RNMSSM.

\subsection{Gluino}
With unbroken $R$ symmetry, 
gauginos such as bino $\tilde{B}$, wino $\tilde{W}$ and gluino $\tilde{g}$ have to be Dirac fermions \cite{Fayet:1978qc,1106.1649} instead of Majorana fermions,
which couple to the fermion components $\tilde{\sigma}_{i}$ in the chiral adjoint superfields $\sigma_i$ as
\begin{eqnarray}{\label{gaugino}}
\mathcal{L}_{\rm{soft}} \supset m_{1}\tilde{\sigma}_{1}\tilde{B}+ m_{2}\tilde{\sigma}_{3}\tilde{W}+m_{3}\tilde{\sigma}_{8}\tilde{g}
\end{eqnarray}
From Eq.(\ref{gaugino}), the R charges of these chiral superfields are found to be zero as shown in Table.\ref{minimal}.
The input values of the gaugino masses $m_i$ in Eq.(\ref{gaugino}) are read from soft SUSY-breaking operators such as
\begin{eqnarray}{\label{goperator}}
\int d^{2}\theta \frac{D_{\alpha}X}{M} W^{\alpha}_{i}\sigma^{i}+\int d^{4}\theta \frac{X^{\dag}X}{M^{2}}\sigma^{\dag}_{i}\sigma_{i},
\end{eqnarray}
where $X$ represents the SUSY-breaking sector\footnote{For discussions about SUSY-breaking sector with $R$ symmetry, see e.g., refs.\cite{0809.1112,Benakli:2008pg, 0812.4600}.} and $M$ denotes the scale at which the SUSY-breaking effect is mediated to the RNMSSM.
Eq.(\ref{goperator}) is consistent with the SM gauge symmetries and the $R$ symmetry as shown in Table.\ref{minimal}, 
where the first and second one yield the one-loop Dirac gaugino masses $m_{i}\sim F/M$
and the  two-loop scalar soft mass squared $m^{2}_{\sigma_{i}}\sim F^{2}/M^{2}$ at the input scale $M$, respectively. 

Just like a Majorana analogy, the Dirac gluino can be directly produced at hadron colliders.
Since the production cross section of the Dirac gluino not only depends on the gluino mass $m_3$ but also squark mass parameters,
we are unable to constrain $m_3$ in model-independent ways. 
For discussions about the constraints from current LHC and future collider experiments, 
see refs.\cite{1111.4322, 1203.4821,1812.09293} and ref.\cite{1606.07090}, respectively.

\subsection{Charginos and Neutralinos}
Without the $SU(2)_L$  triplet $T$ as shown in Table.\ref{minimal},
 the determinant of the charged chargino mass matrix vanishes due to the unbroken $R$ symmetry.
This mass issue can be resolved by an extension of the matter content,
of which the simplest way is the addition of the triplet $T$ \cite{0206102}  which couples to the Higgs doublets as
\begin{eqnarray}{\label{T}}
\delta W=\int d^{2}\theta ~y_{T}H_{d}TH_{u}.
\end{eqnarray}
Eq.(\ref{T}) suggests that $T$ has R charge $+2$.
With T's quantum numbers, one can infer  its soft masses from SUSY-breaking operators 
\begin{eqnarray}{\label{Toperator}}
 \int d^{4}\theta \frac{X^{\dag}X}{M^{2}} T^{\dag}T+ \int d^{2}\theta \frac{D^{2}X}{M}\rm{tr}\left(T\sigma_{3}\right),
\end{eqnarray}
where the first and second term gives rise to the triplet scalar mass squared $m^{2}_{T}\sim F^{2}/M^{2}$  
and  a new Dirac mass parameter $m'_{2}$ similar to $m_2$ in Eq.(\ref{gaugino}), respectively.

\begin{figure*}
\centering
\includegraphics[width=5.5cm,height=7cm]{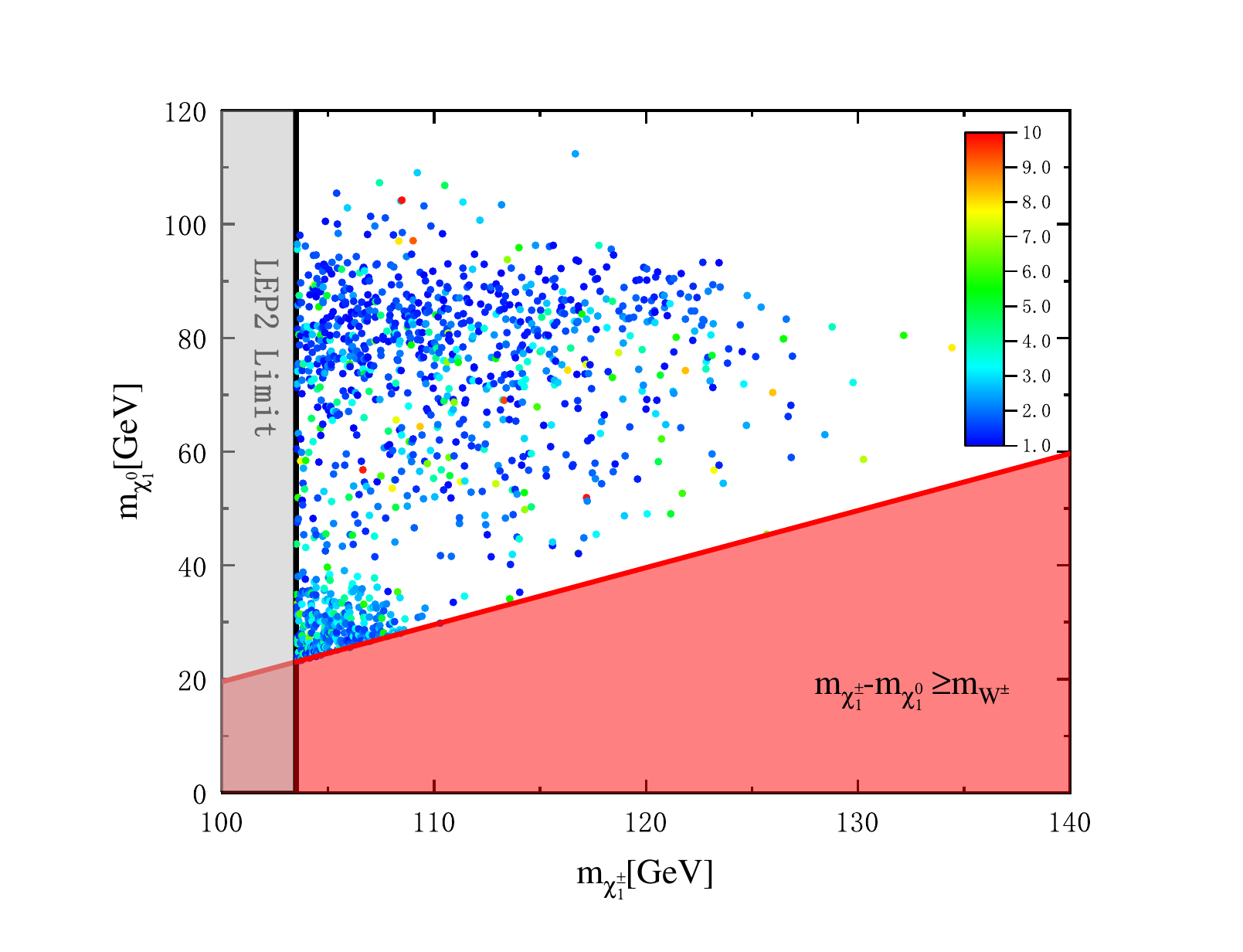}~
\includegraphics[width=5.5cm,height=7cm]{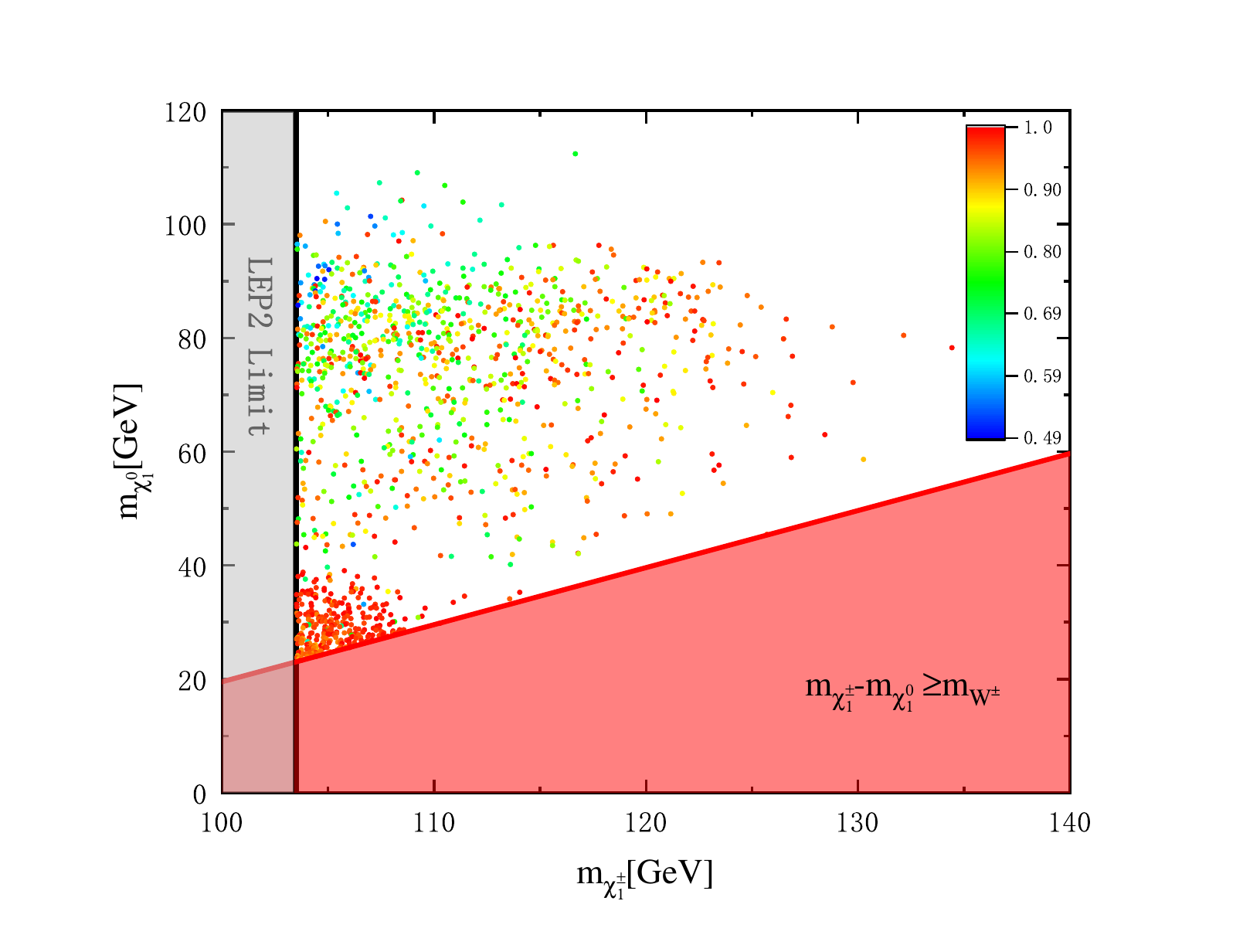}~
\includegraphics[width=5.5cm,height=7cm]{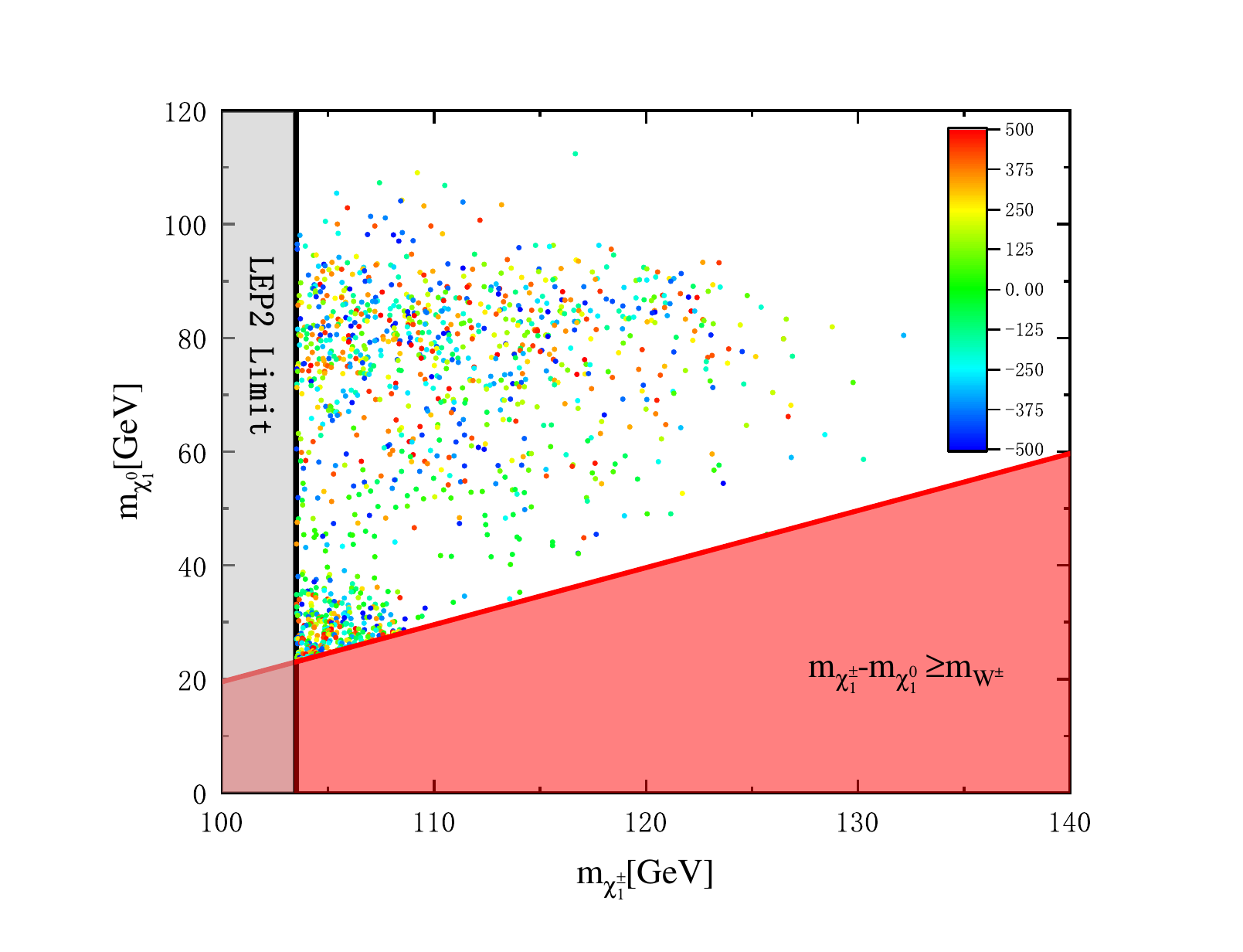}
\centering
\caption{Mass spectrum of $m_{\chi^{\pm}_{1}}$ and $m_{\chi^{0}_{1}}$
with $m_{\chi^{0}_{1}}< m_{\chi^{\pm}_{1}} <m_{\chi^{0}_{1}}+m_{W}$ in the parameter ranges as shown in Table.\ref{ps}, where the dependences of the mass spectrum on $\tan\beta$, $y_T$ and $m_1$ are highlighted in the left, middle and right plot, respectively. 
Regions excluded by the LEP-2 limit (in gray) and disfavored by current LHC data (in red) are shown simultaneously.}
\label{nc}
\end{figure*}

In order to explicitly write the Dirac chargino and neutralino mass matrix, 
it is convenient to divide the triplet fields $\sigma_{3}$ in Eq.(\ref{gaugino}) and $T$ in Eq.(\ref{T})  into the neutral fields $\sigma^{0}_{3}$ and $T^{0}$
and the charged fields $\sigma^{\pm}_{3}$ and $T^{\pm}$ respectively.
With the new superpotential term in Eq.(\ref{T}) taken into account, 
the chargino and neutralino mass matrix are now given by respectively,

\begin{eqnarray}
\left(\begin{array}{cccc} 
\tilde{\sigma}^{-}_{3} & \tilde{T}^{-} & \tilde{W}^{-} & \tilde{H}_{d}^{-}\end{array}\right)
\left(
\begin{array}{cccc}
0& m'_{2} & m_{2}  & 0 \\
m'_{2}  &  0 & 0 & y_{T}\upsilon_{d} \\
m_{2} & 0 & 0  & \sqrt{2}m_{W}\sin\beta \\
 0  &  -y_{T}\upsilon_{u}  &  \sqrt{2}m_{W}\cos\beta & 0  \\
\end{array}\right)
\left(\begin{array}{c}
\tilde{\sigma}^{+}_{3}\\
\tilde{T}^{+}\\
\tilde{W}^{+} \\
\tilde{H}_{u}^{+} \\
\end{array}\right) \label{cmass}
\end{eqnarray}

\begin{eqnarray}
\left(\begin{array}{cccc} 
 \tilde{\sigma}^{0}_{1} & \tilde{\sigma}^{0}_{3} & \tilde{H}_{u}^{0} & \tilde{H}_{d}^{0}\end{array}\right)
\left(
\begin{array}{cccc}
m_{1 }& 0 & 0  & 0\\
0  &  m_{2} & 0 & m'_{2}  \\
m_{Z}\sin\theta_{W} \cos\beta& -m_{Z}\cos\theta_{W} \sin\beta  & -\lambda\upsilon_{d}  & -y_{T}\upsilon_{d}\\
-m_{Z}\sin\theta_{W} \cos\beta   & m_{Z}\cos\theta_{W} \cos\beta  &  -\lambda\upsilon_{u} &   -y_{T}\upsilon_{u}  \\
\end{array}\right)
\left(\begin{array}{c}
\tilde{B}^{0}\\
\tilde{W}^{0} \\
\tilde{N}^{0} \\
\tilde{T}^{0} \\
\end{array}\right) \label{nmass}
\end{eqnarray}

where $\theta_W$ is the weak mixing angle,  the vevs of the two Higgs doublets satisfy $\upsilon^{2}_{u}+\upsilon^{2}_{d}=\upsilon^{2}=(174~\rm{GeV})^{2}$ with $\tan\beta=\upsilon_{u}/\upsilon_{d}$,
and we have simply assumed the singlet vevs
\begin{eqnarray}{\label{vevs}}
\left<\sigma^{0}_{1}\right>=\left<\sigma^{0}_{3}\right>=0.
\end{eqnarray}
Although nonzero $\left<\sigma^{0}_{1}\right>$ and $\left<\sigma^{0}_{3}\right>$ are allowed by the $R$ symmetry,
they make the analysis on the electroweak symmetry breaking 
more complicated than what we have chosen.

In the Dirac chargino mass spectrum given by Eq.(\ref{cmass}),
the lightest mass $m_{\chi^{\pm}_{1}}$ is sensitive to the triplet $T$ 
in the sense that it vanishes as  $T$ disappears, i.e, $y_{T}\rightarrow 0$ and $m'_{2}\rightarrow 0$.
This has to be avoided, 
because $m_{\chi^{\pm}_{1}}$ has to be larger than $\sim 103.5$ GeV from previous LEP-2 limit \cite{Abdallah:2003xe,Abbiendi:2003sc}.
Improved bounds on $m_{\chi^{\pm}_{1}}$ inferred from current LHC data \cite{Aaboud:2018sua, Aad:2014vma}, 
are sensitive to the $\chi^{\pm}_{1}$ decay pattern, as many choices exist due to a wide range of the lightest neutralino mass $m_{\chi^{0}_{1}}$ in Eq.(\ref{nmass}).
In the case $\chi^{\pm}_{1}\rightarrow \chi^{0}_{1}+W^{\pm}$,  $m_{\chi^{\pm}_{1}}$ is excluded up to $600-700$ GeV for massless $\chi^{0}_{1}$ \cite{Aaboud:2018sua}.
If the mass difference between $m_{\chi^{\pm}_{1}}$ and $m_{\chi^{0}_{1}}$ is small,
the $\chi^{\pm}_{1}$  decay is either dominated by $\chi^{\pm}_{1}\rightarrow \chi^{0}_{1}W^{*}\rightarrow \chi^{0}_{1} f\bar{f}'$ or $\chi^{\pm}_{1}\rightarrow W\tilde{G}$, where $f$ and $f'$ refer to SM fermions.
In the former situation the chargino mass bound is greatly relaxed,
while in the later case  $m_{\chi^{\pm}_{1}}$ has to be larger than $\sim 200$ GeV  \cite{Alvarado:2018rfl} for a light gravitino with mass of order $\sim 10$ eV as the lightest supersymmetric particle (LSP).
To summarize,  as $m_{\chi^{\pm}_{1}}$ in our model is less than $\sim 200$ GeV due to a moderate value of $y_T$,
the small mass splitting between $m_{\chi^{\pm}_{1}}$ and $m_{\chi^{0}_{1}}$, 
with $\chi^{0}_{1}$ as the LSP, is favored in light of current collider limits.

\begin{table}
\begin{center}
\begin{tabular}{cc}
\hline\hline
$\rm{parameter}$~~~ & $\rm{range}$ \\ \hline
$m_1$~~~ &  $[-490.8, 490.3]$ \\
$m_2$~~~ &  $[-137.6, 157.1]$   \\
$m'_2$~~~ &  $[-430, 409]$  \\
$\tan\beta$~~~ &  $(1, 10]$ \\
$\lambda$~~~ &  $[0.5, 0.7]$ \\ 
$y_T$~~~ &  $[0.49, 1.0]$ \\
\hline \hline
\end{tabular}
\caption{Parameters ranges in the Dirac chargino-neutralino sector which are consistent with the LEP-2 bound on $m_{\chi^{\pm}_{1}}$ and favored by current LHC data, where masses are in unit of GeV.}
\label{ps}
\end{center}
\end{table}

To illustrate the viability of $m_{\chi^{0}_{1}}< m_{\chi^{\pm}_{1}} <m_{\chi^{0}_{1}}+m_{W}$ in Eqs.(\ref{cmass}) and (\ref{nmass}),
we show in Fig.\ref{nc} the samples which yield the required mass spectrum with the parameter ranges as shown in Table.\ref{ps}, in comparison with regions excluded by the LEP-2 limit (in gray) and disfavored by current LHC data (in red).
We have shown the explicit dependences of the mass spectrum on the model parameters $\tan\beta$ (left), $y_T$ (middle) and $m_1$ (right), respectively.
It indicates that small $m_{\chi^{0}_{1}}$ of order less than $\sim 40$ GeV favors both large $y_{T}\geq 0.9$ and small $\tan\beta \leq 5$,
while moderate $m_{\chi^{0}_{1}}$ of order $\sim 80-100$ GeV has no specific preferences. 
In particular, the right plot in Fig.\ref{nc} implies that a bino-like neutralino dark matter (with small $\mid m_1\mid$) can be achieved in the case of either small or moderate $m_{\chi^{0}_{1}}$. 

Now we turn to the subject of Dirac neutralino dark matter $\chi^{0}_1$.
Since $\chi^{0}_1$ in Eq.(\ref{nmass}) can be either bino- or singlino-like, 
we discuss them separately. 
If $\chi^{0}_1$ is singlino-like, 
the annihilation cross section of $\chi^{0}_1$ pair is mainly through an exchange of Higgs and 
of heavier neutralino states in Eq.(\ref{nmass}).
As $m_{\chi^{0}_{1}}$ in Fig.\ref{nc} is beneath the Higgs mass $m_h$, 
the later annihilation channel is kinetically forbidden.
Such Higgs-portal dark matter \cite{deSimone:2014pda} has been excluded 
by a combination of the Higgs invisible decay in the mass region $m_{\chi^{0}_{1}}<m_{h}/2$ 
and spin-independent direct detection limits in the mass region $m_{\chi^{0}_{1}}>m_{h}/2$,
except a narrow resonant region $m_{\chi^{0}_{1}}\approx m_{h}/2$.

Unlike a Majorana neutralino or the Higgs-portal dark matter as above,
the annihilation cross sections of bino-like $\chi^{0}_1$ (namely the $\tilde{B}^0-\tilde{\sigma}_{1}^{0}$ state) pair into SM fermions is no longer suppressed by the SM fermion masses \cite{0808.2410, 0905.1043, 0911.5273,1307.3561},
which makes the lightest Dirac neutralino a natural realization of leptophilic dark matter \cite{0211325}.
In the situation with squark masses larger than slepton masses, the annihilation of $\chi^{0}_1$ pair is dominated by lepton final states \cite{1307.3561}
\begin{eqnarray}{\label{bino}}
\sigma_{\ell\bar{\ell}}\upsilon_{\rm{rel}}\approx 
\frac{g'^{4}m^{2}_{\chi^{0}_{1}}}{8\pi}\left[\frac{1}{16(m^{2}_{\chi^{0}_{1}}+m^{2}_{\tilde{\ell}_{L}})^{2}}+
\frac{1}{(m^{2}_{\chi^{0}_{1}}+m^{2}_{\tilde{\ell}_{R}})^{2}}\right]\nonumber\\
\end{eqnarray}
where $g'$ is $U(1)_Y$ gauge coupling constant, 
$\upsilon_{\rm{rel}}$ is dark matter relative velocity, 
$\ell=\{\tau,\mu, e\}$, and $m_{\tilde{l}_{L}}$ and $m_{\tilde{l}_{R}}$ refer to the left- and right-hand slepton masses respectively. Note, $R$ symmetry forbids mixing effects between the left- and right-hand sleptons.

\begin{figure}[htb!]
\centering
\includegraphics[width=9cm,height=9cm]{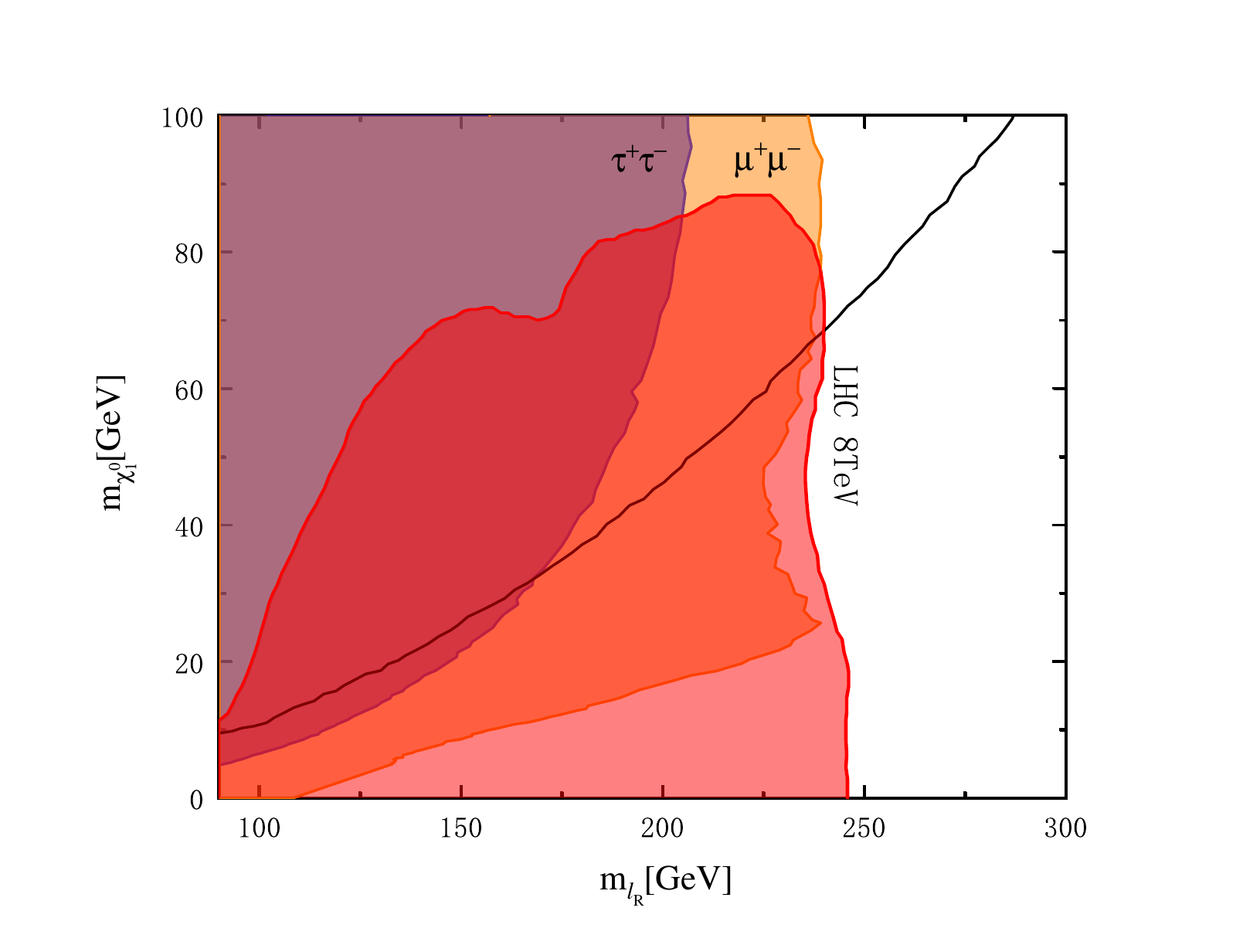}
\centering
\caption{Contour (in black) of  bino-like neutralino dark matter is projected to the two-parameter plane of $m_{\chi^{0}_1}$ and the right-hand slepton masses $m_{\tilde{\ell}_{R}}$ with $\ell=\{\mu,\tau\}$ for selectron masses $m_{\tilde{e}_{L}}=m_{\tilde{e}_{R}}=300$ GeV and left-hand smuon and stau masses $m_{\tilde{\mu}_{L}}=m_{\tilde{\tau}_{L}}=400$ GeV,
in comparison with regions excluded by 8-TeV LHC limit (in red), the AMS limits on $\sigma_{e^{+}e^{-}}\upsilon_{\rm{rel}}$ and $\sigma_{\mu^{+}\mu^{-}}\upsilon_{\rm{rel}}$ (in orange) 
and the Fermi-LAT limit on $\sigma_{\tau^{+}\tau^{-}}\upsilon_{\rm{rel}}$ (in purple).}
\label{bdm}
\end{figure}

Fig.\ref{bdm} shows the contour  (in black)  of the bino-like $\chi^{0}_{1}$ projected to the two-parameter plane of $m_{\chi^{0}_{1}}$ and  the right-hand smuon and stau mass $m_{\tilde{\ell}_{R}}$ with $\ell=\{\mu,\tau\}$, 
which satisfies the dark matter annihilation cross section $\sigma\upsilon_{\rm{rel}}\approx 3\times 10^{-26}\rm{cm}^{3}\rm{s}^{-1}$ from the observed thermal dark matter relic density.
We have taken the selectron masses $m_{\tilde{e}_{L}}=m_{\tilde{e}_{R}}=300$ GeV  consistent with 
the AMS limit on $\sigma_{e^{+}e^{-}}\upsilon_{\rm{rel}}$ \cite{Bergstrom:2013jra} in the mass range $m_{\chi^{0}_{1}}\leq 100$ GeV,
and adopted the left-hand smuon and stau masses $m_{\tilde{\mu}_{L}}=m_{\tilde{\tau}_{L}}=400$ GeV
in the light of LEP \cite{Amsler:2008zzb} and current LHC \cite{Aad:2012pxa, Aad:2014vma} bounds.
\footnote{These mass bounds are mainly based on the slepton pair productions at the LEP and LHC,
which are dominated by the electroweak processes via s-channel $\gamma$, $Z$ or $W$. 
Unlike the Dirac gluino, charginos or neutralinos pair productions \cite{Heikinheimo:2011fk, Choi:2008pi, Choi:2010gc}
which obviously differ from those of Majorana analogies due to new fermion freedoms, 
the slepton pair productions  are only altered due to sub-dominant processes.}
This figure shows that in comparison with the LHC bounds
and the AMS limits on $\sigma_{\mu^{+}\mu^{-}}\upsilon_{\rm{rel}}$ and $\sigma_{\tau^{+}\tau^{-}}\upsilon_{\rm{rel}}$ \cite{Bergstrom:2013jra, Ackermann:2015zua}
$m_{\chi^{0}_{1}}$ between $70$ GeV and $100$ GeV survives with $m_{\tilde{\ell}_{R}}\geq 250$ GeV.

\begin{figure*}
\centering
\includegraphics[width=5.5cm,height=7cm]{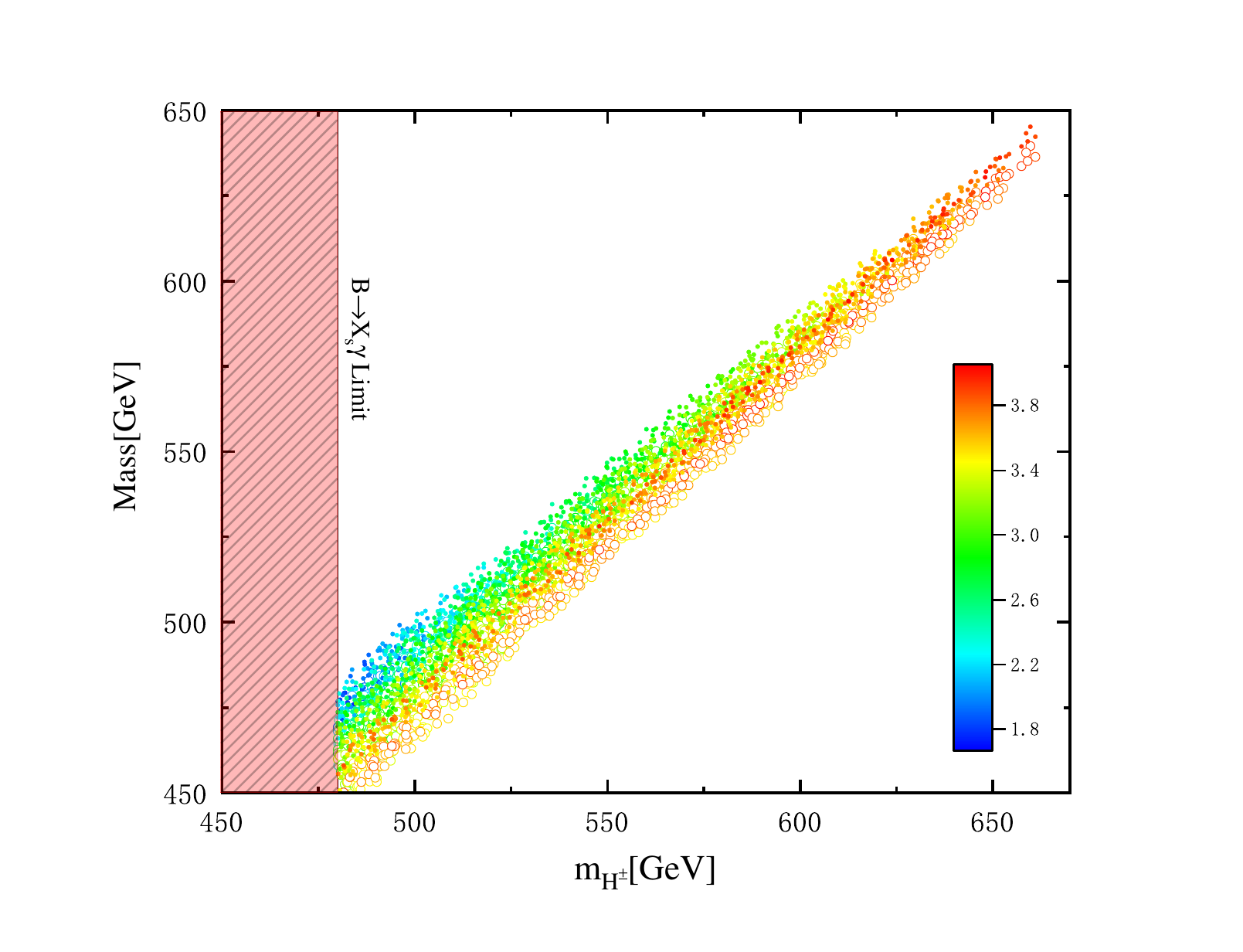}~
\includegraphics[width=5.5cm,height=7cm]{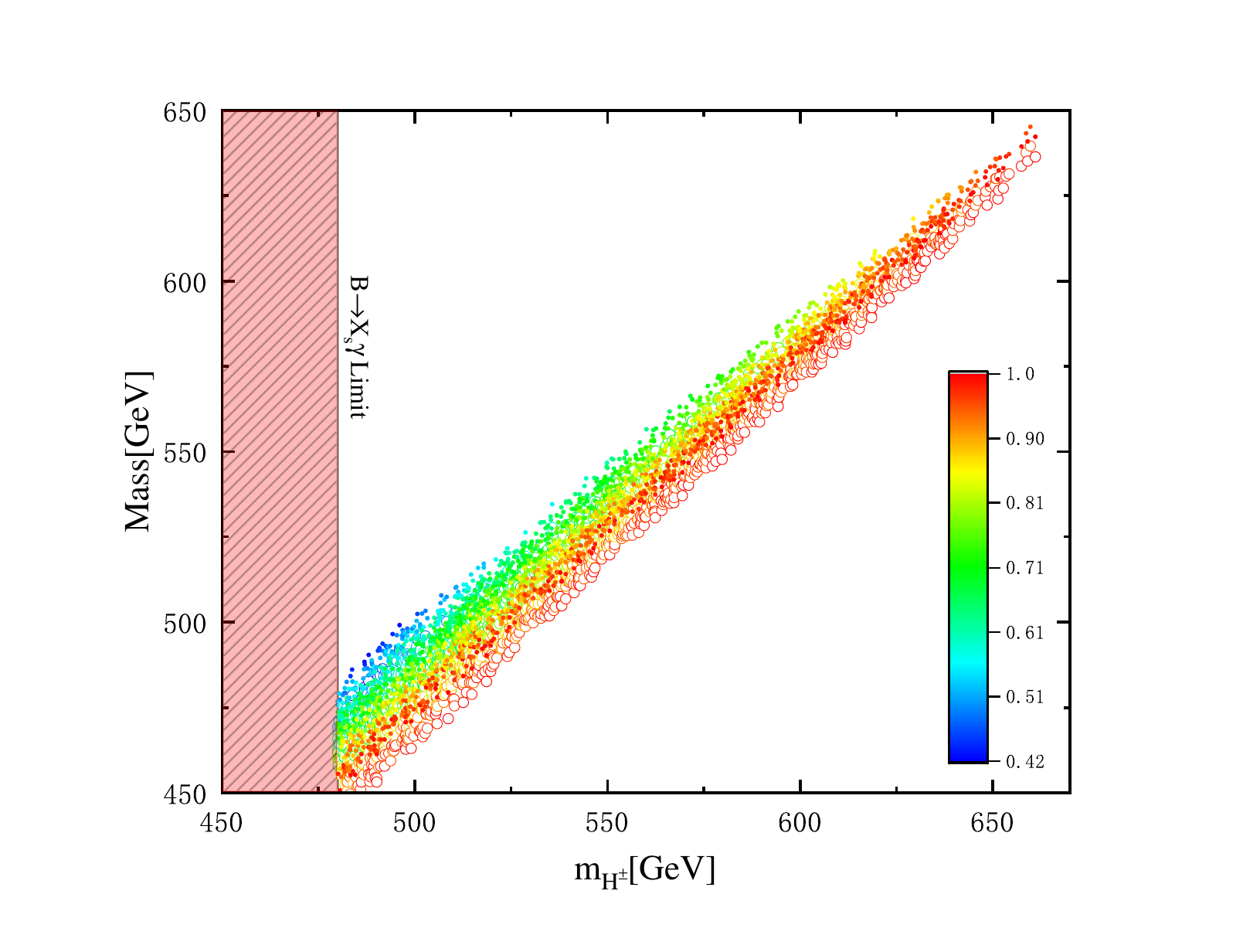}~
\includegraphics[width=5.5cm,height=7cm]{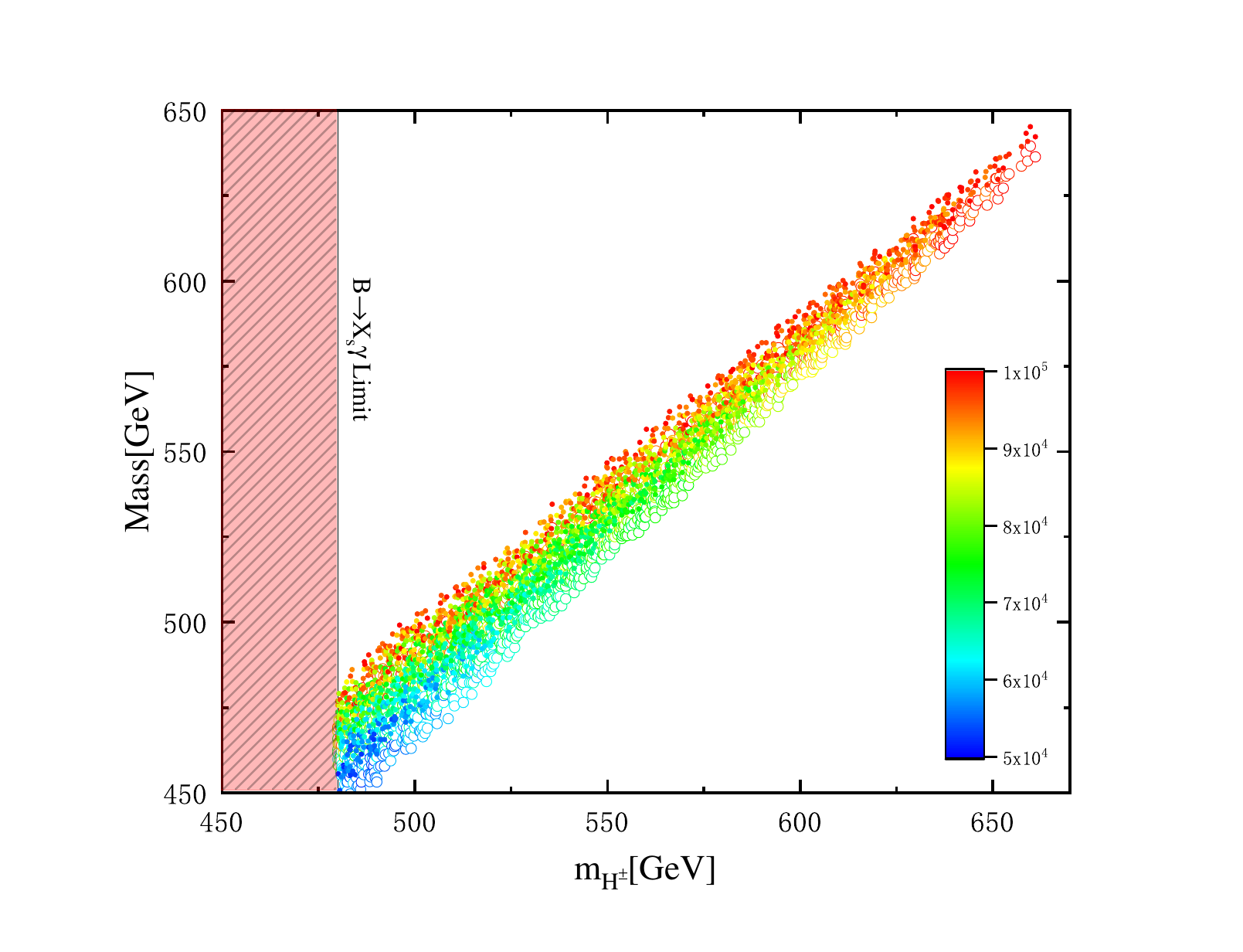}
\centering
\caption{Masses $m_{h_{2}}$ (hollow) and $m_A$ (solid) as function of the CP-charged scalar mass $m_{H^{\pm}}$ with $m_{h_{1}}=125$ GeV in the parameter ranges as shown in Table.\ref{sp},
where the explicit dependences of the mass spectrum on $\tan\beta$, $y_T$ and $b$ are shown in the left, middle and right plot, respectively.}
\label{Hmass}
\end{figure*}

\subsection{Higgs}
Compared to R-symmetric MSSM (RMSSM) \cite{Choi:2010an,Benakli:2012cy, Bertuzzo:2014bwa, Diessner:2014ksa,Diessner:2015iln} or conventional NMSSM, 
the scalar mass spectrum of the Higgs sector in the minimal RNMSSM is different.
Let us begin with relevant soft masses in our model
\begin{eqnarray}{\label{sscalar}}
\mathcal{L}_{\rm{soft}}\supset m^{2}_{H_{u}}H^{\dag}_{u}H_{u}+m^{2}_{H_{d}}H^{\dag}_{d}H_{d}+m^{2}_{N}N^{\dag}N+b H_{u}H_{d}
+ m^{2}_{T}T^{\dag}T,
\end{eqnarray}
where the first three terms arise from the following soft SUSY-breaking operators
\begin{eqnarray}{\label{soperator}}
\int d^{4}\theta \frac{X^{\dag}X}{M^{2}}\left[N^{\dag}N+H_{u}^{\dag}H_{u}+H_{d}^{\dag}H_{d}\right].
 \end{eqnarray}
In Eq.(\ref{soperator}), the two-loop $m^{2}_{N}\sim m^{2}_{H_{u}}\sim m^{2}_{H_{d}} \sim  F^{2}/M^{2}$  are produced at the input scale $M$.
Combing the gauge interactions in Eq.(\ref{Kahler}), the $F$ terms in Eq.(\ref{W}) and the soft masses in Eq.(\ref{sscalar})
gives rise to the scalar mass spectrum in this sector as follows.

Firstly, the mass squared matrixes for the CP-even scalars with $R$ charges $0$ and $2$ are given by 
\begin{eqnarray}
\left(\begin{array}{cc} 
H^{0}_{uR} & H^{0}_{dR} \end{array}\right)
\left(
\begin{array}{cc}
m_{Z}^{2}\cos^{2}\beta+b\tan\beta & [(\lambda^{2}+y_{T}^{2})\upsilon^{2}-\frac{m^{2}_{Z}}{2}]\sin2\beta-b  \\
* &  m_{Z}^{2}\sin^{2}\beta+b\cot\beta  \\
\end{array}\right)
\left(\begin{array}{c}
H^{0}_{uR}\\
H^{0}_{dR} \\
\end{array}\right)\label{smass1}
\end{eqnarray}
and
\begin{eqnarray}
\left(\begin{array}{cc} 
N_{R} & T^{0}_{R} \end{array}\right)
\left(
\begin{array}{cc}
 m^{2}_{N}+\lambda^{2}\upsilon^{2} &  \lambda y_{T}\upsilon^{2} \\
 * &  m^{2}_{T}+m'^{2}_{2}+y_{T}^{2}\upsilon^{2}  \\
\end{array}\right)
\left(\begin{array}{c}
N_{R}\\
T^{0}_{R} \\
\end{array}\right)\label{smass2}
\nonumber\\
\end{eqnarray}
respectively, where the soft masses $m^{2}_{H_{u}}$ and $m^{2}_{H_{d}}$ have been eliminated by the $b$ term in terms of the conditions of electroweak symmetry breaking.

Secondly, the mass squared matrixes for the CP-odd scalars are
\begin{eqnarray}
\left(\begin{array}{cc} 
H^{0}_{uI} & H^{0}_{dI} \end{array}\right)
\left(
\begin{array}{cc}
 b\tan\beta & b  \\
* &  b\cot\beta \\
\end{array}\right)
\left(\begin{array}{c}
H^{0}_{uI} \\
H^{0}_{dI} \\
\end{array}\right)\label{smass3}
\end{eqnarray}
which contains a massless Goldstone mode, and 
\begin{eqnarray}
\left(\begin{array}{cc} 
N_{I} & T^{0}_{I} \end{array}\right)
\left(
\begin{array}{cc}
 m^{2}_{N}+\lambda^{2}\upsilon^{2} &   y_{T}\lambda\upsilon^{2} \\
 * &  m^{2}_{T}+m'^{2}_{2}+y_{T}^{2}\upsilon^{2}  \\
\end{array}\right)
\left(\begin{array}{c}
N_{I} \\
T^{0}_{I} \\
\end{array}\right)\label{smass4}
\nonumber\\
\end{eqnarray}

Thirdly, the mass squared matrix for the CP-charged scalars under the basis $(H^{+}_{u}, H^{-*}_{d})$ is given by
\begin{eqnarray}{\label{cpcharge}}
\mathcal{M}^{2}_{H^{\pm}}=\left[b+\left(\frac{g^{2}_{2}}{2}+y^{2}_{T}-\lambda^{2}\right)\upsilon_{u}\upsilon_{d}\right]
\left(\begin{array}{cc}
\cot\beta &  1 \\
1  &  \tan\beta \\
\end{array}%
\right)\nonumber\\
\end{eqnarray}
with $m^{2}_{T^{\pm}}\approx m^{2}_{T}+m'^{2}_{2}+y^{2}_{T}\upsilon^{2}$, where $g_{2}$ is the  $SU(2)_L$ gauge coupling constant.
From Eq.(\ref{cpcharge}) one obtains a massless Goldstone mode and a massive charged state $H^{\pm}$.
Compared to the NMSSM, 
the new $y_T$ term in Eq.(\ref{cpcharge}) is very useful in uplifting the charged Higgs scalar mass.

The scalars in Eqs.(\ref{smass1})-(\ref{smass4}) can be classified into two sets, 
with one set similar to the type-II Higgs doublets in the MSSM and 
the other set controlled by the free parameters such as $m_N$ and $m_{T}$.
In the former set, the mass relations are however altered,
which at the tree level rely on four input parameters $b$, $\lambda$, $y_T$ and $\tan\beta$.
One can replace the dimensional parameter $b$ by imposing the constraint $m_{h_{1}}=125$ GeV, 
and then uncover the other masses such as $m_{h_{2}}$, $m_A$ and $m_{H^{\pm}}$ by adjusting the three dimensionless parameters.

\begin{table}[htb!]
\begin{center}
\begin{tabular}{cc}
\hline\hline
$\rm{parameter}$~~~ & $\rm{range}$ \\ \hline
$b$~~~ &  $[5\times 10^{4}, 10^{5}]$ $\rm{GeV}^{2}$ \\
$\tan\beta$~~~ &  $[1.6, 4.0]$ \\
$\lambda$~~~ &  $[0.5, 0.7]$ \\ 
$y_{T}$~~~ &  $[0.42, 1.0]$   \\
\hline \hline
\end{tabular}
\caption{Adopted parameters ranges for the numerical analysis in Fig.\ref{Hmass}.}
\label{sp}
\end{center}
\end{table}

So far, the most stringent constraint on these scalars arises from the precision measurements on $B\rightarrow X_{s}\gamma$ \cite{Flacher:2008zq,Misiak:2015xwa, Misiak:2017bgg}.
Fig.\ref{Hmass} shows the masses $m_{h_{2}}$ (hollow) and $m_A$ (solid) as function of the CP-charged scalar mass $m_{H^{\pm}}$ for the parameter ranges as shown in Table.\ref{sp} which satisfy the observed Higgs mass and excess the conservative mass bound $m_{H^{\pm}}\geq 480$ GeV 
 \cite{Misiak:2015xwa} simultaneously.\footnote{The main contribution to $B\rightarrow X_{s}\gamma$ is dominated by the charged Higgs loop diagram,
from which the inferred lower mass bound on $m_{H^{\pm}}$ in the context of type-II Higgs doublet model is valid regardless of whether $R$ symmetry is broken or not.}
The explicit dependences of the mass spectrum on $\tan\beta$, $y_T$ and $b$ are shown in the left, middle and right plot, respectively. 
The lower mass bound implies that both $m_{h_{2}}$ and $m_{A}$ have to exceed $\sim 450$ GeV.
For future prospect of detection on these scalars at the LHC, see e.g., ref.\cite{Akeroyd:2016ymd}.
Note, Eq.(\ref{smass1}) is a tree-level estimate on the observed Higgs mass $m_{h_{1}}$.
It is valid when loop correction is small, which is a natural assumption in our case.

Let us now estimate the fine tuning related to the natural argument of Higgs mass. 
We take the measure of the fine tuning as $\Delta=\max\{\Delta_P\}$ with 
\begin{eqnarray}{\label{tuning}}
\Delta_{P}=\bigg|\frac{\partial \ln m^{2}_{Z}}{\partial \ln P}\bigg|,
\end{eqnarray}
where the soft mass squared $P=\{m^{2}_{i}, b, \cdots\}$.
Among the aforementioned soft masses, 
the dimensional $b$ in Table.\ref{sp} contributes to $\Delta_{b}\sim \tan\beta(b/ 4m^{2}_{Z}) \leq 12.5$.

\subsection{Sfermions}
Apart from Majorana gaugino masses, 
the $R$ symmetry also prohibits holomorphic soft masses such as $A$ terms related to sfermions.
It only allows scalar soft masses
\begin{eqnarray}{\label{scalar}}
\mathcal{L}_{\rm{soft}}\supset m^{2}_{\tilde{f}} f^{\dag}f,~~ \rm{f}=Q,\bar{u},\cdots
\end{eqnarray}
which arise from soft SUSY-breaking operators such as 
\begin{eqnarray}{\label{roperator}}
\int d^{4}\theta \frac{X^{\dag}X}{M^{2}}f^{\dag}f.
\end{eqnarray}
From Eq.(\ref{roperator}), the two-loop $m^{2}_{\tilde{f}}\sim F^{2}/M^{2}$ is generated at the scale $M$.

The absence of $A$ terms  earns us a few advantages.
Naively, the vanishing $A$ term associated with the top quark suggests that the observed Higgs mass cannot be explained without a violation of naturalness,
which is true for typical RMSSM but doesn't stand in the context of RNMSSM.
Because just like the NMSSM there exists large tree-level correction to the Higgs mass in this model.
Moreover, flavor violations, which are too large to violate stringent experimental bounds on the MSSM with a large top-related $A$ terms, are naturally small in the RNMSSM.

\begin{table*}
\centering
\begin{tabular}{ccccc}
\hline\hline
$\rm{Model}$~ & $\rm{Higgs~naturalness~problem}$ & $\mu~\rm{problem}$  &  $\rm{flavor~violation}$ &  $\rm{neutralino~dark~matter}$ \\ \hline
$\rm{MSSM}$ ~ &  $\rm{yes}$ &   $\rm{yes}$   & $\rm{yes}$  & $\rm{Majorana}$  \\
$\rm{NMSSM}$ ~&   $\rm{no}$  &  $\rm{no}$    &  $\rm{yes}$  & $\rm{Majorana}$ \\
$\rm{RMSSM}$ ~&  $\rm{yes}$  &   $\rm{no}$&  $\rm{no}$   &    $\rm{Dirac}$  \\
$\rm{RNMSSM}$~&  $\rm{no}$  &   $\rm{no}$  &  $\rm{no}$  &  $\rm{Dirac}$ \\
\hline \hline
\end{tabular}
\caption{Summary of the main phenomenological features in the RNMSSM compared to the MSSM, the NMSSM and the RMSSM.}
\label{summary}
\end{table*}

\section{Conclusions}
In this study we have proposed a new variant of NMSSM by imposing an unbroken $R$ symmetry.
Compared to the well-known SUSY models such as the MSSM, NMSSM and RMSSM,
the RNMSSM introduces distinctive phenomenological features as shown in Table.\ref{summary}.
Theses features make it a well-motivated scenario that deserves investigation.
We have identified the minimal version of the RNMSSM from the principles of both simplicity and viability. 
Our example, which contains an SM singlet, two $SU(2)_L$ triplets and an $SU(3)_c$ octet beyond the matter content of NMSSM,
indicates that 
\begin{itemize}
\item The $\mu$ problem is resolved by identifying the neutralinos as Dirac fermions. In this model two $SU(2)_L$ triplets have been introduced to illustrate the viability of the idea. 
\item The lightest neutralino mass is typically less than $\sim 100$ GeV.
It can realize the bino-like dark matter with the main annihilations dominated by $\tau^{+}\tau^{-}$ and $\mu^{+}\mu^{-}$.
\item The Higgs mass naturalness is resolved in terms of both the singlet and the triplet which simultaneously contribute to the tree-level Higgs mass.
Meanwhile, they help the charged Higgs mass exceed the lower bound value from the stringent $B\rightarrow X_{s}\gamma$ limit.
\item Due to the absence of $A$ terms flavor violations are naturally small.
\end{itemize}
Finally, although not addressed here, 
there is no obstacle to find an explicit SUSY-breaking sector which yields the soft Lagrangian $\mathcal{L}_{\rm{soft}}$ as we have shown.

\section*{Acknowledgments}
This research is supported in part by the National Natural Science Foundation of China under Grant No.11775039 
and the Fundamental Research Funds for the Central Universities at Chongqing University under Grant No.cqu2017hbrc1B05.

\end{document}